\documentclass[aps,prl,twocolumn,superscriptaddress,showpacs]{revtex4}

\usepackage{graphicx}

\begin{document}

\title{Parametric control of a superconducting flux qubit}

\author{S. Saito}
\affiliation{NTT Basic Research Laboratories, NTT Corporation, 
Atsugi, Kanagawa 243-0198, Japan}
\affiliation{CREST, Japan Science and Technology Agency, Kawaguchi, 
Saitama 332-0012, Japan}

\author{T. Meno}
\affiliation{NTT Advanced Technology Corporation, Atsugi, Kanagawa 243-0198, 
Japan}

\author{M. Ueda}
\affiliation{NTT Basic Research Laboratories, NTT Corporation, 
Atsugi, Kanagawa 243-0198, Japan}
\affiliation{CREST, Japan Science and Technology Agency, Kawaguchi, 
Saitama 332-0012, Japan}
\affiliation{Department of Physics, Tokyo Institute of Technology, 
Meguro, Tokyo 152-8551, Japan}

\author{H. Tanaka}
\affiliation{NTT Basic Research Laboratories, NTT Corporation, 
Atsugi, Kanagawa 243-0198, Japan}
\affiliation{CREST, Japan Science and Technology Agency, Kawaguchi, 
Saitama 332-0012, Japan}

\author{K. Semba}
\affiliation{NTT Basic Research Laboratories, NTT Corporation, 
Atsugi, Kanagawa 243-0198, Japan}
\affiliation{CREST, Japan Science and Technology Agency, Kawaguchi, 
Saitama 332-0012, Japan}

\author{H. Takayanagi}
\affiliation{NTT Basic Research Laboratories, NTT Corporation, 
Atsugi, Kanagawa 243-0198, Japan}
\affiliation{CREST, Japan Science and Technology Agency, Kawaguchi, 
Saitama 332-0012, Japan}

\begin{abstract}

Parametric control of a superconducting flux qubit has been achieved by using 
two-frequency microwave pulses. We have observed Rabi oscillations stemming 
from parametric transitions between the qubit states when the sum of the two 
microwave frequencies or the difference between them matches the qubit Larmor 
frequency. We have also observed multi-photon Rabi oscillations corresponding 
to one- to four-photon resonances by applying single-frequency microwave 
pulses. The parametric control demonstrated in this work widens the frequency 
range of microwaves for controlling the qubit and offers a high quality 
testing ground for exploring nonlinear quantum phenomena. 

\end{abstract}

\pacs{74.50.+r, 03.67.Lx, 42.50.Hz, 85.25.Dq}

\maketitle

Quantum state engineering has become one of the most important arenas in 
quantum physics. In particular, the coherent control of quantum two-state 
systems (TSS), which are applicable to quantum bits (qubit), has attracted 
increasing interest in the context of quantum computing and quantum 
information processing \cite{Nielsen}. Various candidate physical systems 
are being studied for the future implementation of qubits. These include 
artificial quantum TSS like superconducting qubits \cite{Nakamura99,
Vion,Yu,Martinis,Mooij,Chiorescu,Kutsuzawa,Ploude} 
as well as 
naturally existing quantum TSS like nuclear spins \cite{Vandersypen,Yusa}. 
Of the many candidates 
that may enable us to realize quantum computation, superconducting qubits 
based on Josephson junctions have gained increasing importance because of 
their potential controllability and scalability.

The coherent control of a single qubit has been demonstrated in many types 
of superconducting circuits, such as charge \cite{Nakamura99}, charge-phase 
\cite{Vion}, phase \cite{Yu,Martinis}, and flux qubits \cite{Chiorescu,
Kutsuzawa,Ploude}. Recently, two-qubit operation has been demonstrated 
in charge \cite{Yamamoto} and phase qubits \cite{McDermott}. In addition to 
qubit operation, the superconducting qubit offers a testing ground for 
exploring interactions between photons and artificial macroscopic objects, 
which we shall refer to as ``atoms". In the weak-driving limit, the 
interaction 
between a single ``atom" and a single microwave photon has been demonstrated 
with a charge qubit, which is strongly coupled to a superconducting 
transmission line resonator \cite{Wallraff04}. In the strong-driving regime, 
superconducting qubits have exhibited nonlinear optical responses: 
multi-photon Rabi oscillations have been observed in a charge qubit by 
using microwave pulses \cite{Nakamura01}, and under continuous microwave 
irradiation, multi-photon absorption has been observed in a phase 
\cite{Wallraff03} and a flux qubit \cite{Saito}.

In this Letter, we describe the parametric control of a superconducting flux 
qubit with two-frequency microwave pulses. We have succeeded in observing 
two-photon Rabi oscillations of the qubit caused by a parametric transition 
when the qubit Larmor frequency matches either the sum of the two microwave 
frequencies or the difference between them. We also show multi-photon Rabi 
oscillations corresponding to one- to four-photon resonances under 
single-frequency microwave pulse irradiation where the qubit Larmor frequency 
was equal to multiples of the microwave frequency. The parametric and 
multi-photon transitions clearly exhibit high nonlinearity due to interactions 
between a single ``atom" and microwaves. In addition, they can be useful when 
designing practical gate operation on superconducting qubits. For example, 
two-photon processes can be used to entangle two qubits as already shown 
in the field of ion trap qubits \cite{Sackett}. Thus, nonlinear effects will 
offer new possibilities for flexible qubit control as well as being of 
fundamental physical interest in the field of superconducting qubits.

\begin{figure}
\includegraphics[width=1.0\linewidth]{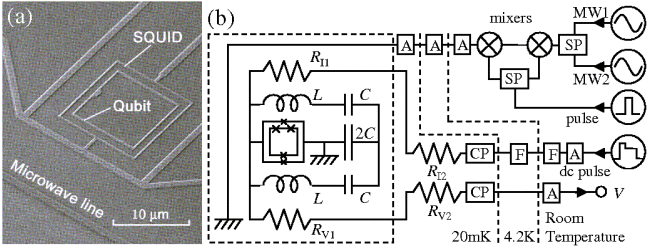}%
\caption{ (a) Scanning electron micrograph of a flux qubit (inner loop) and 
a dc-SQUID (outer loop). 
The loop sizes of the qubit and SQUID are 10.2 $\times$ 10.4~$\mu$m$^{2}$ and 
12.6 $\times$ 13.5~$\mu$m$^{2}$, respectively. They are magnetically 
coupled by the 
mutual inductance  $M\approx$ 13 pH. 
(b) A circuit diagram of the flux qubit measurement system. On-chip 
components are shown in the dashed box. $L\approx$ 140~pH, $C\approx$~9.7 pF, 
$R_{\mathrm{I}1}=$ 0.9~k$\Omega$, $R_{\mathrm{V}1}=$~5 k$\Omega$. Surface 
mount resistors $R_{\mathrm{I}2}=$ 1 k$\Omega$ and $R_{\mathrm{V}2=}$ 
3~k$\Omega$ are set in the sample holder. We put adequate copper powder 
filters CP and LC filters F and attenuators A for each line. 
\label{Fig1}}
\end{figure}

If we are to observe highly nonlinear phenomena in superconducting qubits 
with strong microwave driving, the electromagnetic environment of the qubits 
must be well controlled because the strong driving easily excites unwanted 
environmental resonances, which destroy the qubit coherence. Our device was 
fabricated using electron beam lithography and shadow evaporation techniques 
defining an inner aluminum loop forming the qubit and an outer loop enclosing 
the dc-SQUID loop used for the readout (Fig.~\ref{Fig1}(a)). 
The inner loop 
contains three Josephson junctions, one with an area $\alpha \simeq 0.8$ 
times smaller than the nominally identical area of the other two with 
critical current $I_{\mathrm{c}}\approx 430$ nA. 
The outer loop comprises two Josephson 
junctions of critical current $\approx 140$ nA. 
We placed an 
on-chip microwave line close to the qubit at a distance of 20 $\mu$m so that 
the qubit could be strongly driven by oscillating magnetic fields, which are 
induced by microwave currents flowing through the line. 
To control the electromagnetic environment 
surrounding the qubit, we put resistors $R_{\mathrm{I1}}$, $R_{\mathrm{V1}}$, 
lead inductances $L$, and shunt capacitors $C$, $2C$ on the chip 
(Fig.~\ref{Fig1}(b)). The resistors damp unwanted resonances, which are 
generated in, for example, leads and parasitic capacitors outside the 
resistors. There are two well-controlled resonance modes that are produced 
in the circuits inside the resistors, namely, the on-chip components close 
to the qubit. One is the dc-SQUID's plasma mode with a frequency of 1.0 GHz, 
which is formed in the two symmetrical loops, each being composed of $L$, $C$, 
$2C$, and the SQUID's Josephson inductance. The other is the harmonic LC 
resonance mode with a frequency of 4.311 GHz, which is produced in the larger 
loop consisting of the two $L$'s and the two $C$'s. In this way, we have 
achieved an artificial TSS in a well-controlled environment. 

The three Josephson junctions of the qubit form a double-well potential 
in the space of the Josephson phase when about half a flux quantum threads 
the qubit loop. We use the two lowest levels in the potential as the qubit 
states, which are well separated from the higher levels. Thus, the qubit is 
described by the Hamiltonian $H_{\mathrm{qb}} = (\hbar/2) (\varepsilon 
\sigma_{z}+\Delta \sigma_{x})$, where $\sigma_{x, z}$ are the Pauli spin 
matrices. The eigenstates of $\sigma_{z}$ describe clockwise and 
counter-clockwise persistent currents in the qubit. The qubit tunnel 
splitting is described by $\hbar \Delta$, and $\hbar \varepsilon = 
2I_{\mathrm{p}} \Phi_{0} (\Phi_{\mathrm{qb}}/\Phi_{0}-1.5)$ is the energy 
imbalance between the two potential wells caused by the externally applied 
magnetic flux threading the qubit loop $\Phi_{\mathrm{qb}}$, where 
$\Phi_{0}=h/2e$ is the flux quantum and 
$I_{\mathrm{p}}=I_{\mathrm{c}} \sqrt{1-(1/2\alpha)^{2}}$ is the magnitude of 
the qubit persistent current when the qubit is in the $\sigma_{z}$ 
eigenstates and $I_{\mathrm{c}}$ is the critical current of the larger 
junctions. The energy difference between the ground state $|\mathrm{g}\rangle$ 
and the first excited state $|\mathrm{e}\rangle$ of the qubit is 
$\hbar \omega_{\mathrm{qb}}=hf_{\mathrm{qb}}=\hbar \sqrt{\varepsilon^{2}
+\Delta^{2}}$. Assuming that the applied microwaves are in coherent states, 
we may describe the qubit under microwave irradiation by the Hamiltonian 
\begin{equation}
H=\frac{\hbar}{2} \left( \varepsilon\sigma_{z}+\Delta\sigma_{x} \right)
+\sum_{k=1}^{l} 2\hbar g_{k} \alpha_{k} \sigma_{z} \cos \omega_{\mathrm{MW}k}
t,
\label{eq1}
\end{equation}
where $l$ is one (two) in the case of a single- (two-) frequency microwave and 
$g_{k}$ is the coupling between the qubit and the $k$-th microwave (MW$k$), 
whose amplitude and frequency are $\alpha_{k}$ and $f_{\mathrm{MW}k}
=\omega_{\mathrm{MW}k}/2\pi$, respectively. Solving the Schr\"{o}dinger 
equation with the Hamiltonian~(\ref{eq1}) without the rotating wave 
approximation, we obtain the time evolution of the 
probability $P_{\mathrm{e}}(t)$ with which we find the qubit in 
$|\mathrm{e}\rangle$. The probability $P_{\mathrm{e}}(t)$ oscillates 
periodically under resonant conditions 
$\omega_{\mathrm{qb}}=n_{1}\omega_{\mathrm{MW}1}$ (for the single-frequency 
microwave) or $\omega_{\mathrm{qb}}=|n_{1}\omega_{\mathrm{MW}1} 
\pm n_{2}\omega_{\mathrm{MW}2}|$ (for the two-frequency microwave), 
where $n_{k}$ is the MW$k$ photon number. When we operate the qubit away 
from its degeneracy point $\varepsilon \neq 0$, the frequency of the 
oscillation is given for the single-frequency microwave irradiation by
\begin{equation}
\Omega_{\mathrm{Rabi}}=\frac{\Delta}{A} J_{n_{1}}\left( A \frac{4g_{1}\alpha_{1}}{\omega_{\mathrm{MW}1}}\right)
\label{eq2}
\end{equation}
and for the two-frequency microwave irradiation by
\begin{equation}
\Omega_{\mathrm{Rabi}}=\frac{\Delta}{A} J_{n_{1}}\left( A \frac{4g_{1}
\alpha_{1}}{\omega_{\mathrm{MW}1}}\right)J_{n_{2}}\left( A \frac{4g_{2}
\alpha_{2}}{\omega_{\mathrm{MW}2}}\right).
\label{eq3}
\end{equation}
Here $J_{n_{k}}$ is the $n_{k}$-th order Bessel function of the first kind 
and $A\equiv\varepsilon/\omega_{\mathrm{qb}}\approx1$. This approximation 
is valid when $\varepsilon>>\Delta$. The dressed atom approach gives results 
similar to Eq.~(\ref{eq2}) \cite{Nakamura01,Tannoudji}.

The measurements were carried out in a dilution refrigerator. The sample was 
mounted in a gold plated copper box that was thermalized to the base 
temperature of 20 mK ($k_{\mathrm{B}}T<<\hbar\omega_{\mathrm{qb}}$). 
To produce two-frequency microwave pulses, we added two 
microwaves MW1 and MW2 with frequencies of $f_{\mathrm{MW}1}$ and 
$f_{\mathrm{MW}2}$, respectively by using a splitter SP (Fig.~\ref{Fig1}(b)). 
Then we shaped them into microwave pulses 
through two mixers. 
We measured the amplitude of MW$k$ $V_{\mathrm{MW}k}$ at the point between 
the attenuator and the mixer with an oscilloscope. 
We confirmed that unwanted 
higher-order frequency components in the pulses, for example 
$|f_{\mathrm{MW}1} \pm f_{\mathrm{MW}2}|$, $2f_{\mathrm{MW}1}$, and 
$2f_{\mathrm{MW}2}$ are negligibly small under our experimental conditions. 
First, we choose the operating point by setting $\Phi_{\mathrm{qb}}$ 
around 1.5$\Phi_{0}$, which fixes the qubit Larmor frequency 
$f_{\mathrm{qb}}$. The qubit is thermally initialized to be in 
$|\mathrm{g} \rangle$ by waiting for 300 $\mu$s, which is much longer than 
the qubit energy relaxation time (for example 3.8 $\mu$s at 
$f_{\mathrm{qb}}=$ 11.1~GHz). Then a qubit operation is performed by applying a microwave pulse to the qubit. The pulse, with an appropriate length 
$t_{\mathrm{p}}$, amplitudes $V_{\mathrm{MW}k}$, and frequencies 
$f_{\mathrm{MW}k}$, prepares a qubit in the superposition state of 
$|\mathrm{g} \rangle$ and $|\mathrm{e} \rangle$. After the operation, 
we immediately apply a dc readout pulse to the dc-SQUID. This dc pulse 
consists of a short (70 ns) initial pulse followed by a long (1.5 $\mu$s) 
trailing plateau that has 0.6 times the amplitude of the initial part. 
For $\Phi_{\mathrm{qb}}<1.5\Phi_{0}$, if the qubit is detected as being in 
$|\mathrm{e} \rangle$, the SQUID switches to a voltage state and an output 
voltage pulse should be observed; otherwise there should be no output voltage 
pulse. By repeating the measurement 8000 times, we obtain the SQUID switching 
probability $P_{\mathrm{sw}}$, which is directly related to 
$P_{\mathrm{e}}(t_{\mathrm{p}})$ for the dc readout pulse with a proper 
amplitude. For $\Phi_{\mathrm{qb}}>1.5\Phi_{0}$, $P_{\mathrm{sw}}$ is 
directly related to $1-P_{\mathrm{e}}(t_{\mathrm{p}})$.

\begin{figure}
\includegraphics[width=1.0\linewidth]{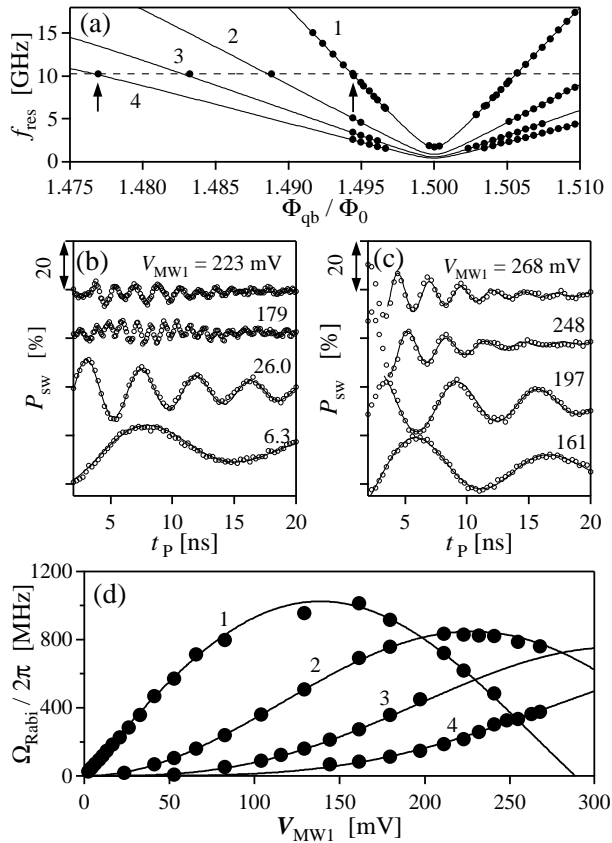}%
\caption{Experimental results with single-frequency microwave pulses. 
(a) Spectroscopic data of the qubit. Each set of 
the dots represents the resonant frequencies $f_{\mathrm{res}}$ 
caused by the one to four-photon absorption processes. 
The solid curves are numerical fits. 
The dashed line shows a microwave frequency 
$f_{\mathrm{MW}1}$ of 10.25~GHz. 
(b) One-photon Rabi oscillations of $P_{\mathrm{sw}}$ with exponentially 
damped oscillation fits. 
Both the qubit Larmor frequency $f_{\mathrm{qb}}$ 
and the microwave frequency $f_{\mathrm{MW}1}$ are 10.25~GHz. The external 
flux is $\Phi_{\mathrm{qb}}/\Phi_{0}$ = 1.4944. 
(c) Four-photon Rabi oscillations when 
$f_{\mathrm{qb}}=$ 41.0~GHz, $f_{\mathrm{MW}1}=$ 10.25~GHz, and 
$\Phi_{\mathrm{qb}}/\Phi_{0}=$ 1.4769. 
(d) The microwave amplitude dependence 
of the Rabi frequencies $\Omega_{\mathrm{Rabi}}/2\pi$ up to four-photon Rabi 
oscillations. The solid curves represent theoretical fits.\label{Fig2}}
\end{figure}

We performed a spectroscopy measurement of the qubit with long (50~ns) 
single-frequency microwave pulses. We observed multi-photon resonant peaks 
($\Phi_{\mathrm{qb}}<1.5\Phi_{0}$) and dips ($\Phi_{\mathrm{qb}}>1.5\Phi_{0}$) 
in the dependence of $P_{\mathrm{sw}}$ on $f_{\mathrm{MW}1}$ at a fixed 
magnetic flux $\Phi_{\mathrm{qb}}$. We obtained the qubit energy diagram 
by plotting their positions as a function of $\Phi_{\mathrm{qb}}/\Phi_{0}$ 
(Fig.~\ref{Fig2}(a)). We took the data around the degeneracy point 
$\Phi_{\mathrm{qb}} \approx 1.5\Phi_{0}$ by applying an additional dc pulse 
to the microwave line to shift $\Phi_{\mathrm{qb}}$ away from $1.5\Phi_{0}$ 
just before the readout, because the dc-SQUID could not distinguish 
the qubit states 
around the degeneracy point. The top solid curve in Fig.~\ref{Fig2}(a) 
represents a numerical fit to the resonant frequencies of one-photon 
absorption. From this fit, we obtain the qubit parameters $E_{\mathrm{J}}/h=$ 
213~GHz, $\Delta/2\pi=$ 1.73~GHz, and $\alpha=$ 0.8. The other curves 
in Fig.~\ref{Fig2}(a) are drawn 
by using these parameters for $n_{1}$ = 2, 3, and 4.

Next, we used short single-frequency microwave pulses with a frequency of 
10.25~GHz to observe the coherent 
quantum dynamics of the qubit. Figures~\ref{Fig2}(b) and (c) show one- and 
four-photon Rabi oscillations observed at the operating points indicated by 
arrows in Fig.~\ref{Fig2}(a) with various microwave amplitudes 
$V_{\mathrm{MW}1}$. 
These data can be fitted by damped oscillations $\propto \exp(-t_{\mathrm{p}}
/T_{\mathrm{d}}) \cos(\Omega_{\mathrm{Rabi}}t_{\mathrm{p}})$, except 
for the upper two curves in Fig.~\ref{Fig2}(b). Here, $t_{\mathrm{p}}$ and 
$T_{\mathrm{d}}$ are the microwave pulse length and qubit decay time, 
respectively. To obtain $\Omega_{\mathrm{Rabi}}$, we performed a fast 
Fourier transform (FFT) on the curves that we could not fit by damped 
oscillations. Although we controlled the qubit environment, there were some 
unexpected resonators coupled to the qubit, which could be excited by the 
strong microwave driving or by the Rabi oscillations of the qubit. We consider 
that these resonators degraded the Rabi oscillations in the higher 
$V_{\mathrm{MW}1}$ range of Fig.~\ref{Fig2}(b). Figure~\ref{Fig2}(d) shows 
the $V_{\mathrm{MW}1}$ dependences of $\Omega_{\mathrm{Rabi}}/2\pi$ up to 
four-photon Rabi oscillations, which are well reproduced by Eq.~(\ref{eq2}). 
Here, we used only one scaling parameter $a$(10.25~GHz) = 0.013 defined as 
$a(f_{\mathrm{MW1}})\equiv 4g_{1}\alpha_{1}/\omega_{\mathrm{MW}1}
V_{\mathrm{MW}1}$, because it is hard to measure the real amplitude of 
the microwave applied to the qubit at the sample position. The scaling 
parameter $a(f_{\mathrm{MW1}})$ reflects the way in which the applied 
microwave is attenuated during its transmission to the qubit and the 
efficiency of the coupling between the qubit and the on-chip microwave line. 
In this way, we can estimate the real microwave amplitude and the interaction 
energy between the qubit and the microwave $2\hbar g_{1}\alpha_{1}$ by fitting 
the dependence of $\Omega_{\mathrm{Rabi}}/2\pi$ on $V_{\mathrm{MW}1}$. 
These results show that we can reach a driving regime that is so strong that 
the interaction energy $2\hbar g_{1}\alpha_{1}$ is larger than the qubit 
transition energy $\hbar \omega_{\mathrm{qb}}$.

We have also performed experiments with two microwave friequencies 
$f_{\mathrm{MW}1}$ and $f_{\mathrm{MW}2}$.
First, we carried out a spectroscopy measurement by using long (50 ns) 
two-frequency microwave pulses. In addition to resonances caused by the 
multi-photon absorption processes at multiples of each microwave frequency 
($f_{\mathrm{qb}}=n_{1}f_{\mathrm{MW}1}$, $n_{2}f_{\mathrm{MW}2}$), 
we also clearly observed those due to parametric processes 
($f_{\mathrm{qb}}=|f_{\mathrm{MW}1} \pm f_{\mathrm{MW}2}|$) (not shown).

\begin{figure}
\includegraphics[width=1.0\linewidth]{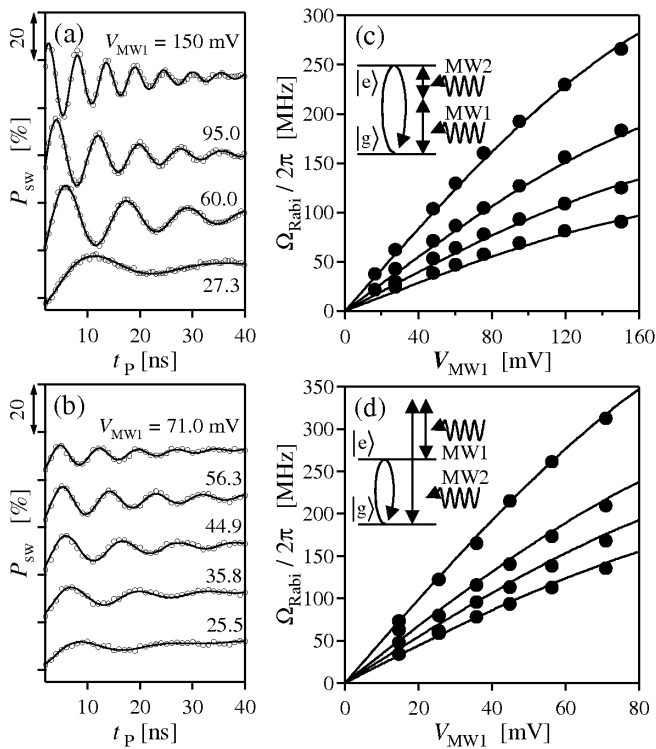}
\caption{Experimental results with two-frequency microwave pulses. 
(a) [(b)] Two-photon Rabi oscillations due to a parametric process 
when $f_{\mathrm{qb}}=f_{\mathrm{MW}2} +[-]\ f_{\mathrm{MW}1}$.
The solid curves are fits by exponentially 
damped oscillations. 
(c) [(d)] Rabi frequencies as a function of $V_{\mathrm{MW}1}$ , 
which are obtained 
from the data in Fig.~\ref{Fig3}(a) [(b)]. The dots represent experimental 
data when $V_{\mathrm{MW}2}=$ 16.9, 23.5, 33.0, and 52.0 
[50.1, 62.9, 79.1, and 124.7]~mV from the bottom set 
of dots to the top one. The solid curves represent Eq.~(\ref{eq3}). 
The inset is a schematic of the parametric process that causes 
two-photon Rabi oscillation 
when $f_{\mathrm{qb}}=f_{\mathrm{MW}2} + [-]\ f_{\mathrm{MW}1}$. 
\label{Fig3}}
\end{figure}

We next investigated the coherent oscillations of the qubit through the 
parametric processes by using short two-frequency microwave pulses. 
Figure~\ref{Fig3}(a) [(b)] shows the Rabi oscillations of $P_{\mathrm{sw}}$ 
when 
the qubit Larmor frequency $f_{\mathrm{qb}}=$ 26.45 [7.4]~GHz corresponds 
to the sum of 
the two microwave frequencies $f_{\mathrm{MW}1}=$ 16.2~GHz, 
$f_{\mathrm{MW}2}=$ 10.25~GHz [the difference between 
$f_{\mathrm{MW}1}=$ 11.1~GHz and $f_{\mathrm{MW}2}=$ 
18.5~GHz] and the
microwave amplitude of MW2 $V_{\mathrm{MW}2}$ was fixed at 33.0 [50.1]~mV. 
They are 
well fitted by exponentially damped oscillations 
$\propto \exp(-t_{\mathrm{p}}/T_{\mathrm{d}}) \cos(\Omega_{\mathrm{Rabi}}
t_{\mathrm{p}})$. 
The Rabi frequencies obtained from the data in Fig.~3(a) [(b)] are well 
reproduced 
by Eq.~(\ref{eq3}) without any fitting parameters (Fig.~\ref{Fig3}(c) [(d)]). 
Here, we used $\Delta$, which was obtained from the spectroscopy measurement 
(Fig.~\ref{Fig2}(a)) and used  $a$(10.25~GHz) = 0.013 
and  $a$(16.2 GHz) = 0.0074 [$a$(11.1~GHz) = 0.013 and 
$a$(18.5~GHz) = 0.0082], which had been obtained from Rabi oscillations 
by using single-frequency microwave pulses with each frequency. Those results 
provide strong evidence that we can achieve parametric control of the qubit 
with two-frequency microwave pulses.

In summary, we investigated nonlinear responses in the superconducting 
flux qubit. 
First, we observed multi-photon Rabi oscillations caused by up to four-photon 
transitions by using single-frequency microwave pulses. The microwave 
amplitude dependences of the Rabi frequencies are well reproduced by Bessel 
functions derived from a semi-classical model. Furthermore, 
we successfully demonstrated parametric control of the qubit by using 
two-frequency microwave pulses. We observed Rabi oscillations of the qubit 
caused by parametric transitions when $f_{\mathrm{qb}}=|f_{\mathrm{MW}1} 
\pm f_{\mathrm{MW}2}|$. The Rabi frequencies as a function of 
the microwave amplitudes are well described by the product of two Bessel 
functions. 
These results indicate that the flux qubit offers 
a good testing ground for exploring quantum nonlinear phenomena 
in a macroscopic quantum object. Furthermore, these multi-photon processes 
observed in our experiment widen the frequency range of microwaves for 
controlling flux qubit.

\begin{acknowledgments}
We thank T. Kutsuzawa, F. Deppe for useful discussions and for helping with 
the experimental setup. We also acknowledge useful discussions with 
J. Johansson, H. Nakano, M. Thorwart, Y. Yamamoto, J. E. Mooij, 
C. J. P. M. Harmans, and Y. Nakamura. This work has been supported by 
the CREST project of the Japan Science and Technology Agency (JST).
\end{acknowledgments}

\end{document}